\documentclass{tcibook}
\usepackage{fancyhea}
\usepackage{work}
\usepackage{bm}       
\usepackage{graphicx}
\usepackage{hyperref}      

\newcommand{\nc}{\newcommand}  

\def\beq{\begin{equation}}
\def\eeq#1{\label{#1}\end{equation}}
\def\eeqn{\end{equation}}

\newenvironment{Eqnarray}%
   {\arraycolsep 0.14em\begin{eqnarray}}{\end{eqnarray}}
\def\beqa{\begin{Eqnarray}}
\def\eeqa#1{\label{#1}\end{Eqnarray}}
\def\eeqan{\end{Eqnarray}}

\nc{\ra}{\rightarrow}  
\nc{\slsh}{\slash\hspace*{-0.22cm}}
\def\Re{{\cal R \mskip-4mu \lower.1ex \hbox{\it e}\,}}
\def\Im{{\cal I \mskip-5mu \lower.1ex \hbox{\it m}\,}}

\nc{\vev}[1]{ \left\langle {#1} \right\rangle }
\nc{\bra}[1]{ \langle {#1} | }
\nc{\ket}[1]{ | {#1} \rangle }
\nc{\fb}{\,{\rm fb}^{-1}}
\nc{\ev}{{\rm eV}}
\nc{\kev}{{\rm keV}}
\nc{\Mev}{{\rm MeV}}
\nc{\gev}{{\rm GeV}}
\nc{\tev}{{\rm TeV}}
\nc{\mev}{{\rm MeV}}

\def\del{\partial}
\def\Dslash{\not{\hbox{\kern-4pt $D$}}}
\def\dslash{\not{\hbox{\kern-2pt $\del$}}}
\def\pslash{\not{\hbox{\kern-2pt $p$}}}
\def\ETmiss{ \not{\hbox{\kern-4pt $E$}}_T }

\def\msb{{\bar{\ssstyle M \kern -1pt S}}}

\begin{document}

\def\bibname{References}
\bibliographystyle{plain}

\raggedbottom

\pagenumbering{roman}

\parindent=0pt
\parskip=8pt
\setlength{\evensidemargin}{0pt}
\setlength{\oddsidemargin}{0pt}
\setlength{\marginparsep}{0.0in}
\setlength{\marginparwidth}{0.0in}
\marginparpush=0pt


\pagenumbering{arabic}

\renewcommand{\chapname}{chap:intro_}
\renewcommand{\chapterdir}{.}
\renewcommand{\arraystretch}{1.25} 
\addtolength{\arraycolsep}{-3pt}


 
\chapter{Z Prime: A Story \\ \textsf{\large{A Boson, A Paper Detector, and a Future Accelerator}}}
\label{chap:Zprime}

\begin{center}\begin{boldmath}



\begin{center}

\begin{large} {\bf D. Hayden, C. Willis, and R. Brock} \end{large}
(Michigan State University).

\end{center}



\end{boldmath}\end{center}


\section{Introduction}
\label{sec:EF-intro}

This is an imaginary story that could come to pass involving a detector, a Boson, and good resolution. 

Vector Bosons in addition to the $W^\pm$ and $Z^0$ have been a theme of particle searches for decades, being a common feature of many models aiming to describe nature beyond the Standard Model (SM).  A new heavy vector boson would likely be one of the first clearly visible signals for new physics to be detected by an experiment, when a new accelerator switches on and/or higher centre of mass energies achieved.  This is due to the resonant production and inherently higher cross-section than other SM background processes at a given polemass, as well as in most models there being a modest branching fraction to very clean decay modes such as leptonic channels.

This story focuses on a possible new vector boson called the $Z^{\prime}$, which in its most basic incarnation, the Sequential Standard Model (SSM)~\cite{TheoryZp}, is depicted as a gauge boson with the same couplings as the SM $Z^0$ but a much higher polemass (on the order of TeV). A group theoretical realization of this model is to add an additional $U(1)^{\prime}$ symmetry to the existing SM structure ($SU(3)_C \times SU(2)_L \times U(1)_Y$). In nearly all models considered (but to varying degrees), this new boson interferes with its SM counterpart leading to an additional degree of deviation from the SM expectation in the mass spectrum preceeding the peak.  Depending on the properties of the new boson, another interesting effect would be in the angular distribution of events in the so-called Collin-Soper frame~\cite{CS}, which could aid a discovery search in certain scenarios, and importantly help distinguish between different signal models for new physics once a discovery is made.

The theoretically-motivated $E_6$ Grand Unified Theory (GUT) model~\cite{TheoryZp, TheoryE6_1, TheoryE6_2}, invokes two extra $U(1)^{\prime}$ symmetries that occur through the decomposition $E_6 \rightarrow SO(10) \times U(1)_{\psi} \rightarrow SU(5) \times U(1)_{\chi} \times U(1)_{\psi}$ (where $SU(5)$ is the gauge group containing the SM suggested by Georgi and Glashow in 1974~\cite{TheorySU5}).  The mixing of these extra $U(1)^{\prime}$ symmetries lead to a new gauge boson: $Z^{\prime}(\theta) = Z^{\prime}_{\psi} \cos\theta + Z^{\prime}_{\chi} \sin\theta$, where the mixing angle $\theta$ determines the coupling to fermions and results in various possible model variations with specific $Z^{\prime}$ states, such as $\theta = 0$ which implies one of the narrowest $E_6$ resonances, $Z^{\prime}_{\psi}$.

Another well motivated and distinct model involving the $Z^{\prime}$ is the Left-Right Symmetric Model (LRM)~\cite{TheoryZp}. This model is derived from a decomposition of the $SO(10)$ GUT, where a right-handed gauge group is added to the electroweak sector of the SM, restoring parity at high energy by replacing $SU(2)_L$ with $SU(2)_L \times SU(2)_R$, and $U(1)_Y$ with $U(1)_{B-L}$.  In the same way that $SU(2)_L \times U(1)_Y$ generates the electroweak sector in the SM, $SU(2)_R \times U(1)_{B-L}$ then gives rise to $W^{\prime\pm}$ and $Z^{\prime}$ additional gauge bosons.

The scenario played out in this paper was to search for both a $Z^{\prime}_{LR}$ and $Z^{\prime}_{\psi}$ decaying to dileptons.  This imagining assumed that nature conspired for a $Z^{\prime}_{LR}$ gauge boson to exist at a polemass of 3~TeV.

\section{The Simulation}

For all Snowmass studies the Delphes-3 fast simulation framework~\cite{Delphes} was used to bring into life a new apparatus called the Snowmass Detector. This future detector was designed to consist of the best and forseeably upgraded components of the current ATLAS (A Toroidal LHC Apparatus)~\cite{ATLAS} and CMS (Compact Muon Solenoid)~\cite{CMS} detectors located at the LHC (Large Hadron Collider) at CERN in Geneva, Switzerland.  The Delphes framework supports the simulation of pile-up events (PU), and parameterizes realistic detector performance and measurements based on full simulation.

Given a newly built Snowmass Detector, the Delphes framework is then capable of producing various accelerator experimental setups and data taking environments.  For this study Delphes was used to simulate proton-proton (pp) collisions at a theoretical future collider with a 14~TeV, and later 33~TeV, centre of mass energy (compared to the LHC's current $\sqrt{s}$ = 8~TeV and maximum 14~TeV capability).  The framework is also capable of simulating events with different pile-up scenarios, namely: 0, 50, and 140 PU.  For this search, each of the pile-up scenarios was investigated, but due to the very clean dilepton final state signature, it was determined that the pile-up scenario had a negligible effect on the result and thus all generated samples were run with PU = 0.

The backgrounds to this search were centrally produced~\cite{Backgrounds} for Snowmass studies, providing an adequate number of generated events to describe the SM processes decaying to dileptons at $\sqrt{s}$ = 14 (33)~TeV for integrated luminosities well over the studied scenarios of 300 and 3000~fb$^{-1}$, up to very high invariant masses. Therefore the SM dilepton background to a 3~TeV resonance is described in Monte Carlo (MC) with a good statistical precision.

\section{Event Selection}

The SM background composition relevant to this search consists of contributions from both reducible and irreducible processes.  Irreducible processes such as Drell-Yan ($q\bar{q} \rightarrow Z/\gamma^* \rightarrow \ell\bar{\ell}$) have many event kinematics which are indistinguishable from the signal process ($q\bar{q} \rightarrow Z^{\prime} \rightarrow \ell\bar{\ell}$), and thus relies on observables such as the differential cross-section of dilepton events ($d \sigma/d m_{\ell\ell}$), which should be steeply falling at high-mass for the SM, but sharply peak at the polemass of the signal process.  The Drell-Yan process represents by far the most dominant background in this search, however it is important to assess other possible sources of signal contamination due to SM processes. Reducible processes are those from the SM that mimic the search signature $\ell\bar{\ell}$ but are inherently different processes with either non-prompt/multiple leptons, or jets which fake a lepton. The reducible SM processes relevant to this search are: $t\bar{t}$, W+jets, multi-jets, and diboson processes such as WW, WZ, and ZZ.  An example of the signal selection contamination from these processes would be W+jets where the W decays leptonicly ($W^+ \rightarrow \ell^+\nu_{\ell}$) and the jet fakes an electron to a sufficient degree that is passes the event selection.  With the variables available in the Monte Carlo, a modest event selection was chosen to preferentially select the signal process and suppress reducible backgrounds.  The event selection criteria requires at least two leptons ($e^+e^-$ or $\mu^+\mu^-$) in the event, with each lepton having $p_T$ $>$ 25~GeV and $|\eta|$ $<$ 2.5.  If more than two pairs of same-flavour opposite-sign leptons pass this criteria, the highest $p_T$ pair is taken, and subjected to the final criterion that the dilepton invariant mass ($m_{\ell\ell}$) be greater than 80~GeV.  If these criteria are met then the event is kept in the analysis, otherwise it is rejected.

\section{The Discovery at the LHC with $\sqrt{s}$ = 14~TeV}

The classic search is to look for peaks in the invariant mass distribution of two oppositely charged, same flavour leptons. Electrons provide a clean, high-resolution observable with backgrounds from essentially all Drell-Yan sources. At masses beyond current limits, there is essentially no complication from pile-up, and resolutions are dominated by the constant term in the resolution function.

Muons are a less-well resolved signal, but an observation at a common invariant mass in both channels would be a striking signal and difficult to argue away on the basis of fluctuation. So while they may not contribute to precision width determination, they would be an essential confirmation, especially at the low integrated luminosities of an early running of the 14~TeV collider.

\subsection{Run 1 of the LHC}

\begin{figure}
\centering
\begin{minipage}{.5\textwidth}
  \centering
  \includegraphics[width=\linewidth]{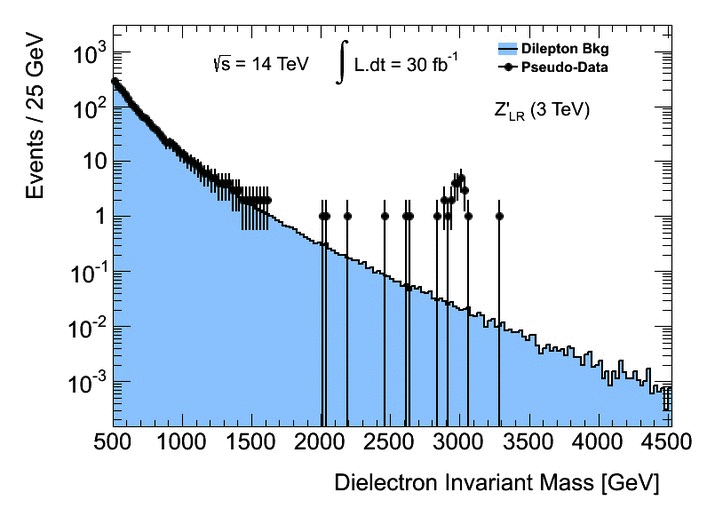}
  \caption{Dilepton backgrounds and the emerging signal for a LR $Z^{\prime}$ at 3~TeV for $e^+e^-$ pairs after 30~fb$^{-1}$.}
  \label{SigPlusBkg_ee_14TeV_30fb}
\end{minipage}%
\begin{minipage}{.5\textwidth}
  \centering
  \includegraphics[width=\linewidth]{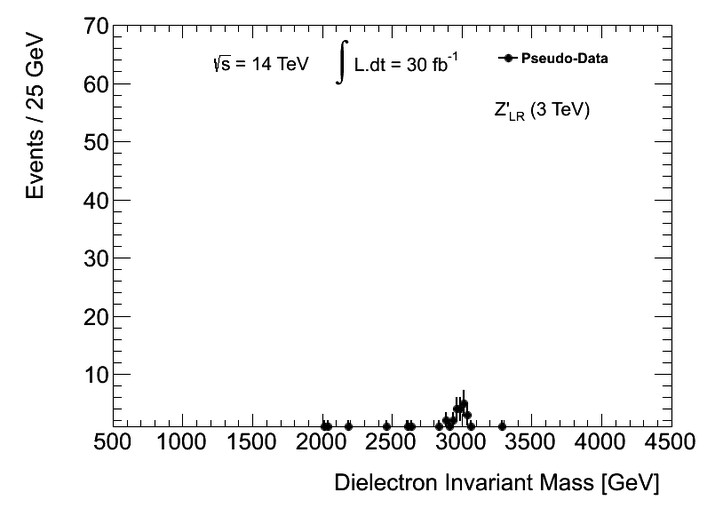}
  \caption{Emerging signal for a LR $Z^{\prime}$ at 3~TeV, background subtracted for $e^+e^-$ pairs after 30~fb$^{-1}$.}
  \label{Sig_ee_14TeV_30fb}
\end{minipage}
\end{figure}

The first 14~TeV run of the LHC started on January 1, 2015 and the Snowmass Detector accumulated an integrated luminosity of 100 fb$^{-1}$ over a period of two years, collecting with a steady rate throughout. By summer of the first year, physicists began to detect a marginal enhancement inconsistent with background in the 3~TeV region (see Figures~\ref{SigPlusBkg_ee_14TeV_30fb} and \ref{Sig_ee_14TeV_30fb}) and continued to watch it as the run progressed.

\begin{figure}
\centering
\begin{minipage}{.5\textwidth}
  \centering
  \includegraphics[width=\linewidth]{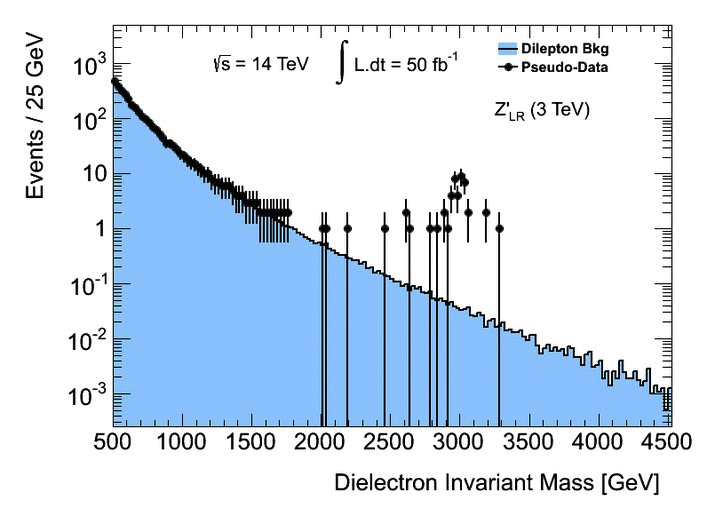}
  \caption{Dilepton backgrounds and the emerging signal for a LR $Z^{\prime}$ at 3~TeV for $e^+e^-$ pairs after 50~fb$^{-1}$.}
  \label{SigPlusBkg_ee_14TeV_50fb}
\end{minipage}%
\begin{minipage}{.5\textwidth}
  \centering
  \includegraphics[width=\linewidth]{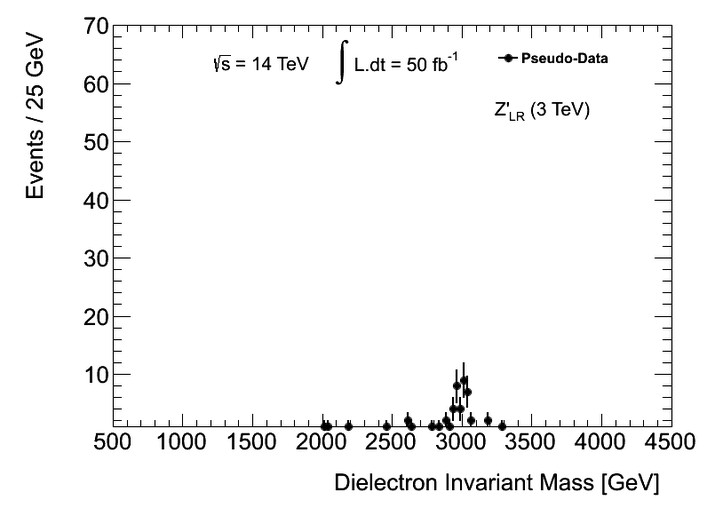}
  \caption{Emerging signal for a LR $Z^{\prime}$ at 3~TeV, background subtracted for $e^+e^-$ pairs after 50~fb$^{-1}$.}
  \label{Sig_ee_14TeV_50fb}
\end{minipage}
\end{figure}

\begin{figure}
\centering
\begin{minipage}{.5\textwidth}
  \centering
  \includegraphics[width=\linewidth]{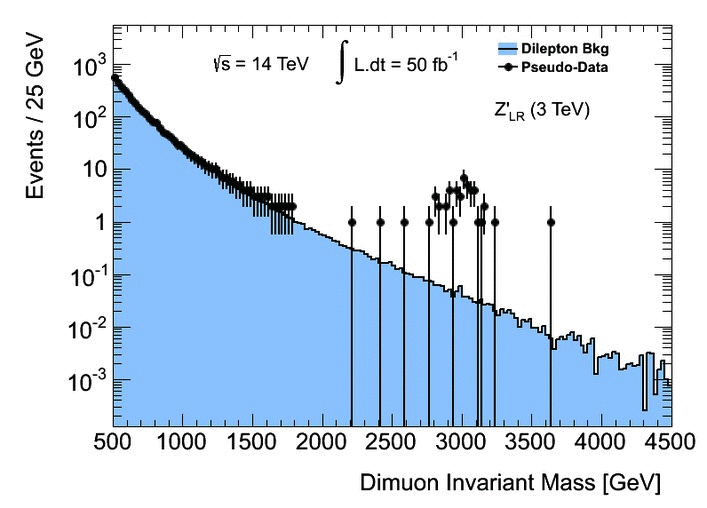}
  \caption{Dilepton backgrounds and the emerging signal for a LR $Z^{\prime}$ at 3~TeV for $\mu^+\mu^-$ pairs after 50~fb$^{-1}$.}
  \label{SigPlusBkg_mm_14TeV_50fb}
\end{minipage}%
\begin{minipage}{.5\textwidth}
  \centering
  \includegraphics[width=\linewidth]{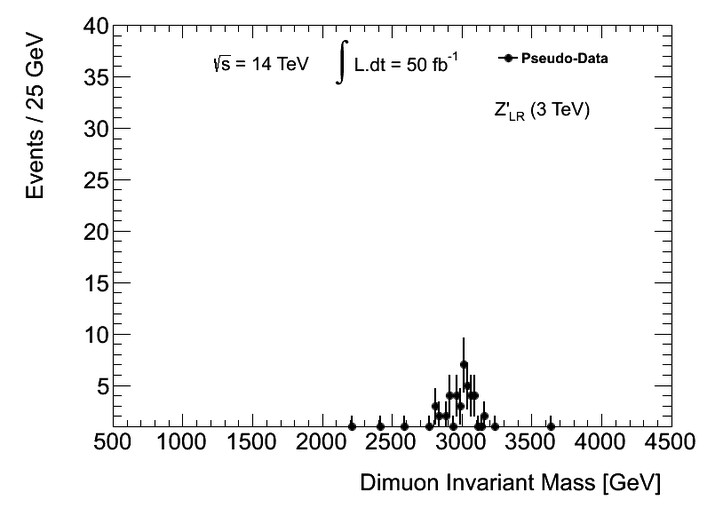}
  \caption{Emerging signal for a LR $Z^{\prime}$ at 3~TeV, background subtracted for $\mu^+\mu^-$ pairs  after 50~fb$^{-1}$.}
  \label{Sig_mm_14TeV_50fb}
\end{minipage}
\end{figure}

The bump was noticeably growing as the summer of 2015 arrived and analyzers began to look seriously at the muon channel with eager anticipation. By this time, 50 fb$^{-1}$ was on disk and the situation had become exciting, classing the observation as evidence for new physics (see Figures~\ref{SigPlusBkg_ee_14TeV_50fb}/\ref{Sig_ee_14TeV_50fb} for the electron channel, and Figures~\ref{SigPlusBkg_mm_14TeV_50fb}/\ref{Sig_mm_14TeV_50fb} for the muon channel).

\begin{figure}
\centering
\begin{minipage}{.5\textwidth}
  \centering
  \includegraphics[width=\linewidth]{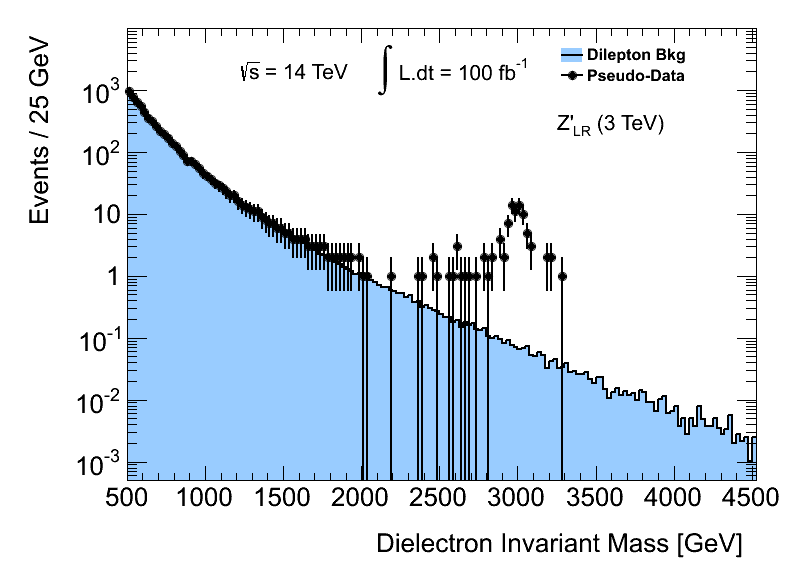}
  \caption{Dilepton backgrounds and the emerging signal for a LR $Z^{\prime}$ at 3~TeV for $e^+e^-$ pairs after 100~fb$^{-1}$.}
  \label{SigPlusBkg_ee_14TeV_100fb}
\end{minipage}%
\begin{minipage}{.5\textwidth}
  \centering
  \includegraphics[width=\linewidth]{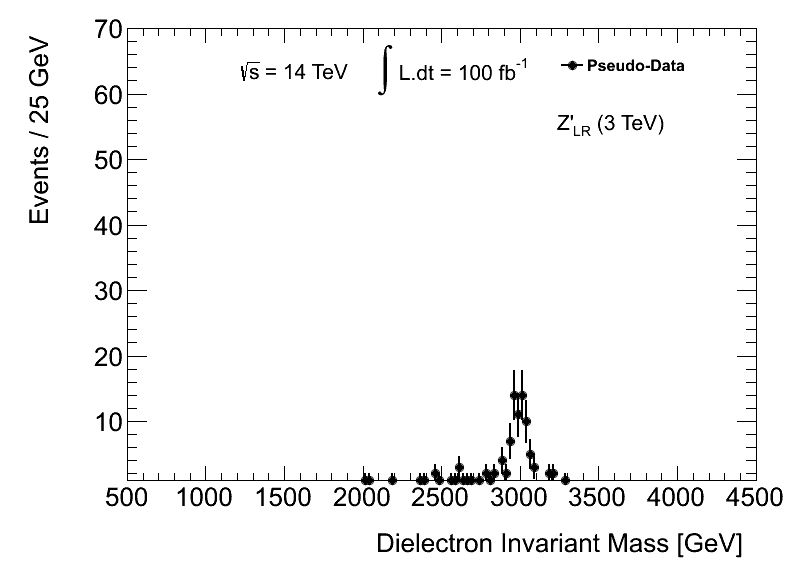}
  \caption{Emerging signal for a LR $Z^{\prime}$ at 3~TeV, background subtracted for $e^+e^-$ pairs after 100~fb$^{-1}$.}
  \label{Sig_ee_14TeV_100fb}
\end{minipage}
\end{figure}

\begin{figure}
\centering
\begin{minipage}{.5\textwidth}
  \centering
  \includegraphics[width=\linewidth]{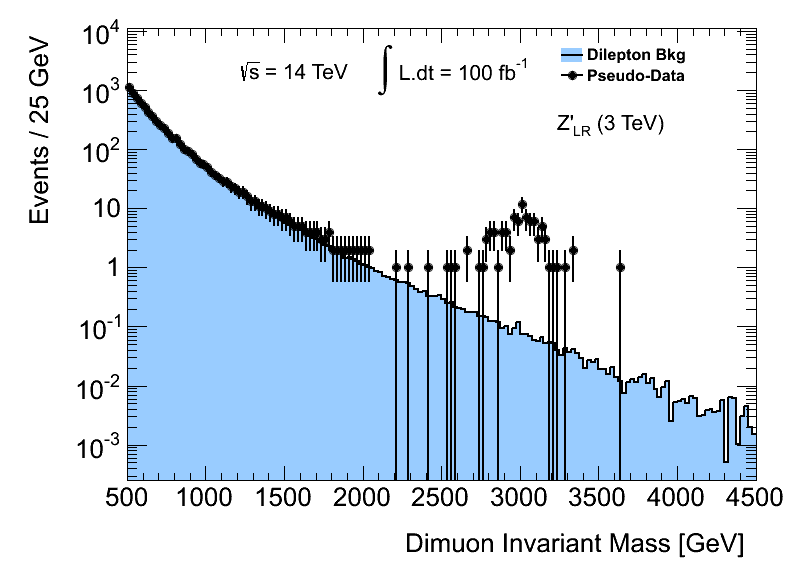}
  \caption{Dilepton backgrounds and the emerging signal for a LR $Z^{\prime}$ at 3~TeV for $\mu^+\mu^-$ pairs after 100~fb$^{-1}$.}
  \label{SigPlusBkg_mm_14TeV_100fb}
\end{minipage}%
\begin{minipage}{.5\textwidth}
  \centering
  \includegraphics[width=\linewidth]{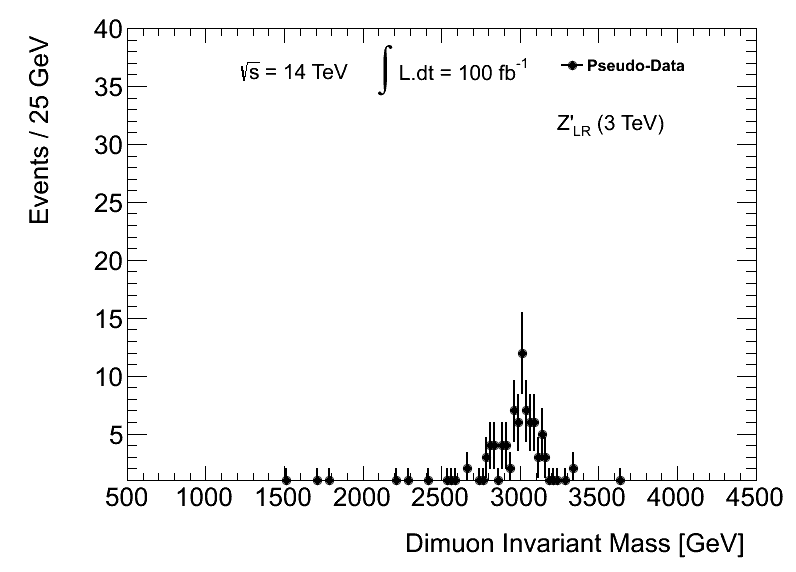}
  \caption{Emerging signal for a LR $Z^{\prime}$ at 3~TeV, background subtracted for $\mu^+\mu^-$ pairs  after 100~fb$^{-1}$.}
  \label{Sig_mm_14TeV_100fb}
\end{minipage}
\end{figure}

Obviously, all eyes were on the Snowmass Experiment as by the end of Run 2 (Figures~\ref{SigPlusBkg_ee_14TeV_100fb} to \ref{Sig_mm_14TeV_100fb}), there were clear peaks in both channels and a discovery had been declared. But what kind of resonant-like new physics had been discovered in the dilepton final state? It was time to upgrade the detectors for the first high luminosity running, and the LHC ceased operations with the expectation that perhaps a few dozen $Z'$ candidates had been produced.

\subsection{Run 2 of the LHC}

\begin{figure}
\centering
\begin{minipage}{.5\textwidth}
  \centering
  \includegraphics[width=\linewidth]{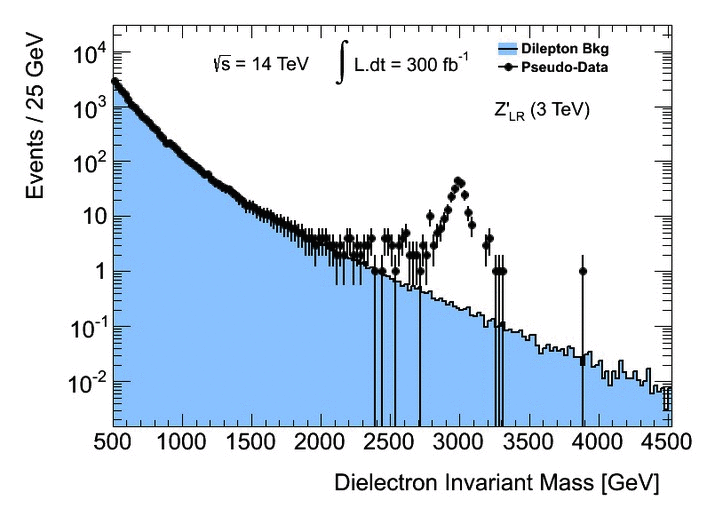}
  \caption{Dilepton backgrounds and the emerging signal for a LR $Z^{\prime}$ at 3~TeV for $e^+e^-$ pairs  after 300~fb$^{-1}$.}
  \label{SigPlusBkg_ee_14TeV_300fb}
\end{minipage}%
\begin{minipage}{.5\textwidth}
  \centering
  \includegraphics[width=\linewidth]{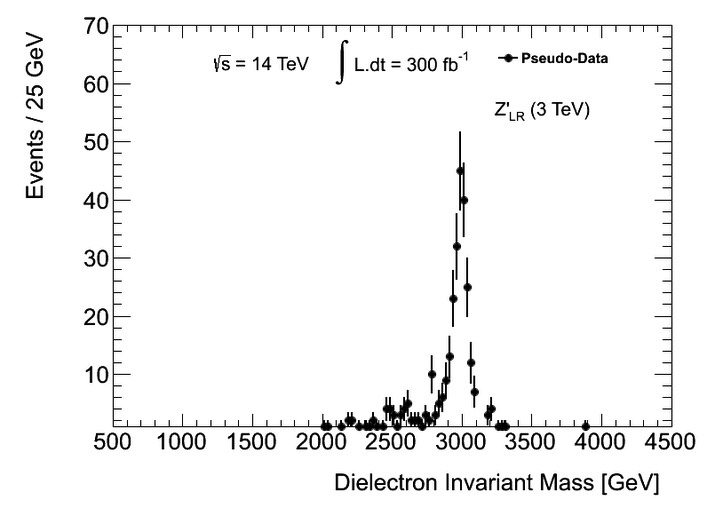}
  \caption{Emerging signal for a LR $Z^{\prime}$ at 3~TeV, background subtracted for $e^+e^-$ pairs after 100~fb$^{-1}$.}
  \label{Sig_ee_14TeV_300fb}
\end{minipage}
\end{figure}

\begin{figure}
\centering
\begin{minipage}{.5\textwidth}
  \centering
  \includegraphics[width=\linewidth]{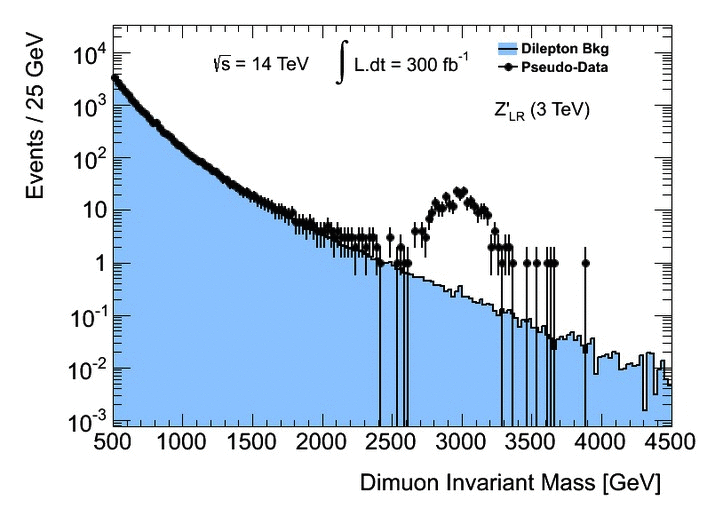}
  \caption{Dilepton backgrounds and the emerging signal for a LR $Z^{\prime}$ at 3~TeV for $\mu^+\mu^-$ pairs after 300~fb$^{-1}$.}
  \label{SigPlusBkg_mm_14TeV_300fb}
\end{minipage}%
\begin{minipage}{.5\textwidth}
  \centering
  \includegraphics[width=\linewidth]{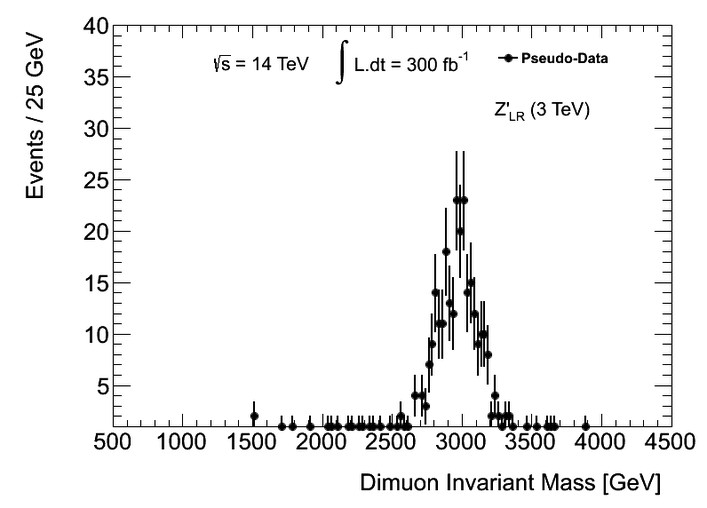}
  \caption{Emerging signal for a LR $Z^{\prime}$ at 3~TeV, background subtracted for $\mu^+\mu^-$ pairs  after 100~fb$^{-1}$.}
  \label{Sig_mm_14TeV_300fb}
\end{minipage}
\end{figure}

Run 2 of the LHC started on January 1, 2019. The accelerator and detector were significantly upgraded and the run started smoothly, but at a higher rate of collisions.  By the end of Run 2, 300~fb$^{-1}$ had been collected by the Snowmass detector (Figures~\ref{SigPlusBkg_ee_14TeV_300fb} to \ref{Sig_mm_14TeV_300fb}), tripling the Run 1 total dataset, which allowed physicists to start to try and determine the nature of the new resonance that had been discovered.

\begin{figure}
\centering
\begin{minipage}{.5\textwidth}
  \centering
  \includegraphics[width=\linewidth]{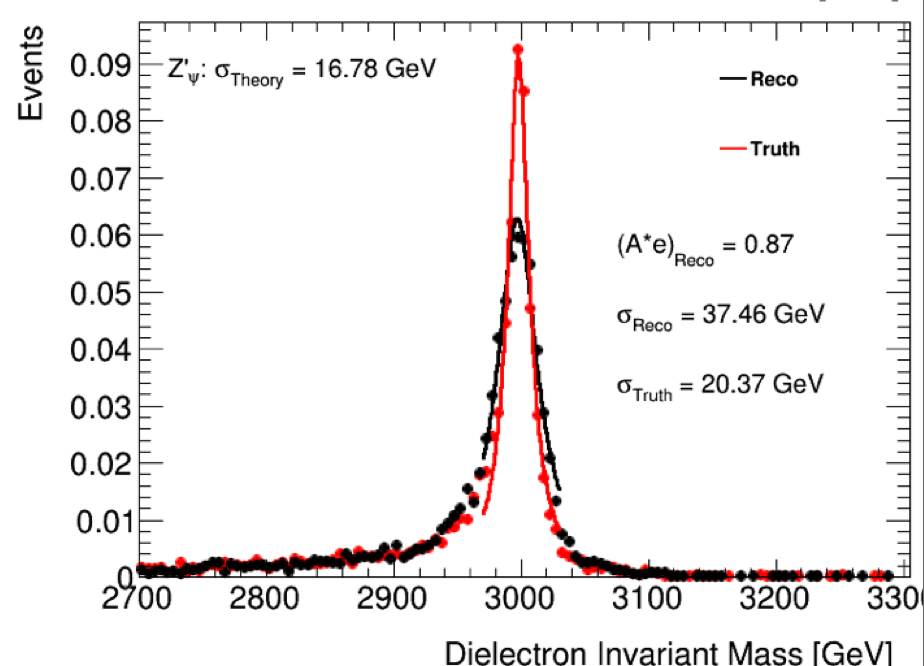}
  \caption{True and Snowmass Detector invariant mass comparison for for a LR $Z^{\prime}$ at 3~TeV, background subtracted for $e^+e^-$ pairs.}
  \label{Resolution_ee_14TeV_300fb}
\end{minipage}%
\begin{minipage}{.5\textwidth}
  \centering
  \includegraphics[width=\linewidth]{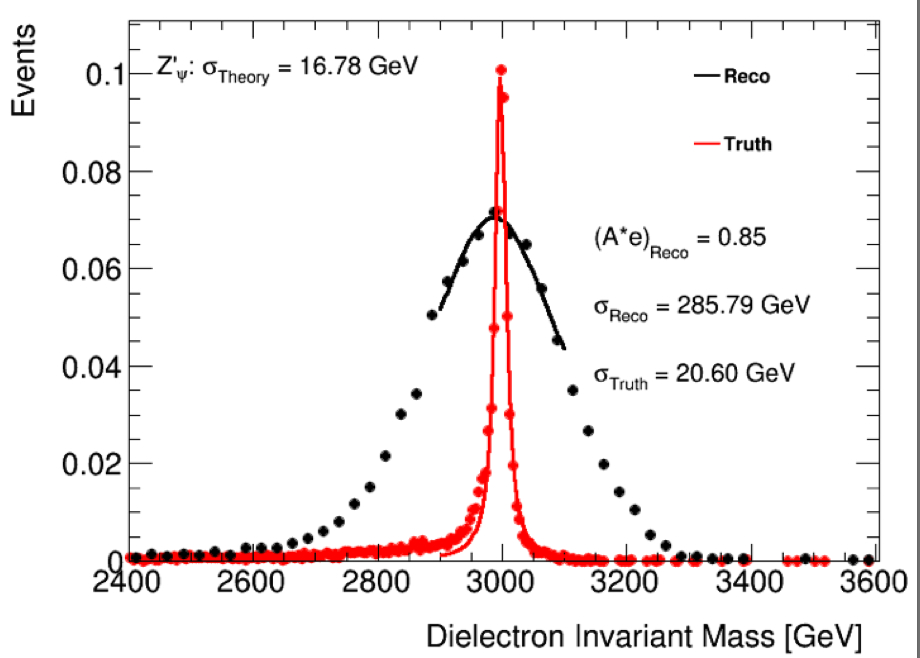}
  \caption{True and Snowmass Detector invariant mass comparison for for a LR $Z^{\prime}$ at 3~TeV, background subtracted for $\mu^+\mu^-$ pairs.}
  \label{Resolution_mm_14TeV_300fb}
\end{minipage}
\end{figure}

\begin{figure}
\centering
\begin{minipage}{.5\textwidth}
  \centering
  \includegraphics[width=\linewidth]{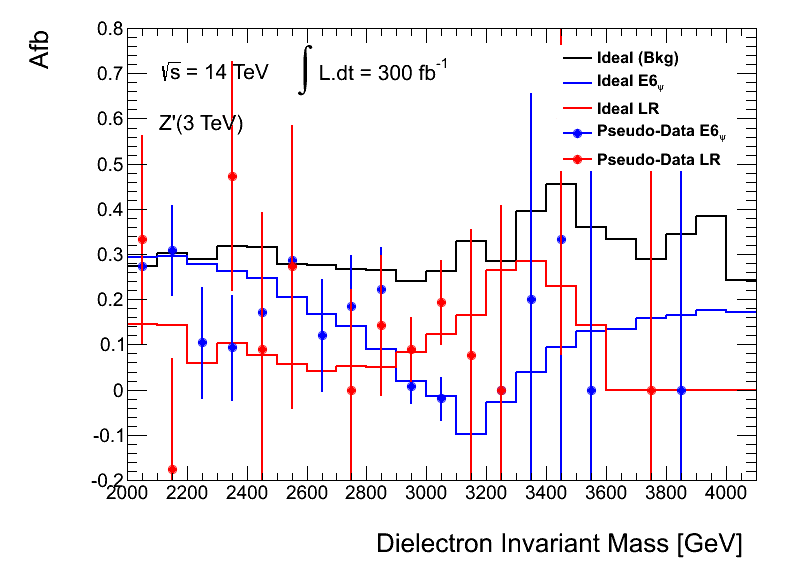}
  \caption{$A_{FB}$ of $e^+e^-$ pairs for the expected SM background (Black), as well as two signal scenarios for a 3~TeV resonance: $E_6$ model $Z^{\prime}_{\psi}$ (Blue), and LR model $Z^{\prime}_{LR}$ (Red). The solid lines show the ideal distributions, and colored data points show a single pseudo experiment after 300~fb$^{-1}$.}
  \label{Afb_ee_14TeV_300fb}
\end{minipage}%
\begin{minipage}{.5\textwidth}
  \centering
  \includegraphics[width=\linewidth]{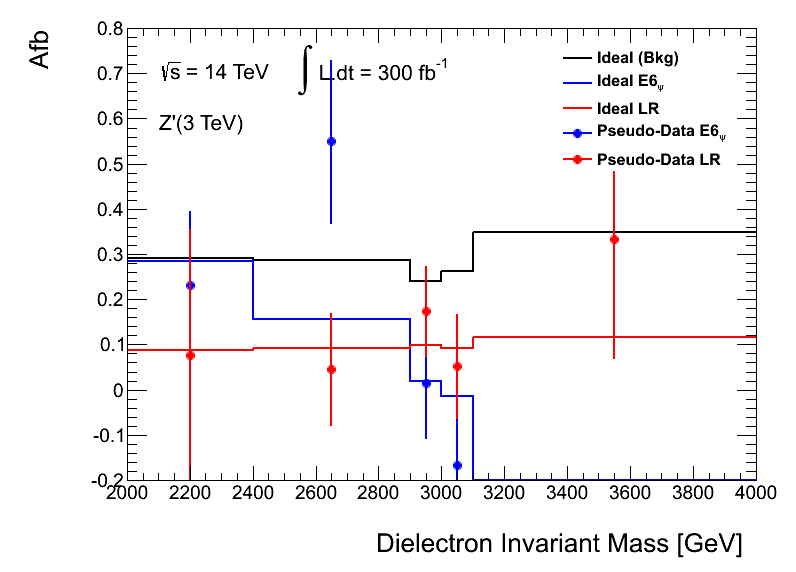}
  \caption{$A_{FB}$ of $e^+e^-$ pairs for the expected SM background (Black), as well as two signal scenarios for a 3~TeV resonance: $E_6$ model $Z^{\prime}_{\psi}$ (Blue), and LR model $Z^{\prime}_{LR}$ (Red). The solid lines show the ideal distributions, and colored data points show a single pseudo experiment after 300~fb$^{-1}$. Here the tail statistics are combined into fewer bins.}
  \label{Afb_ee_14TeV_300fb_Bin}
\end{minipage}
\end{figure}

The energy and mass resolution of the Snowmass Detector was well understood, comparing experiences the physcists had with that of past experiments such as ATLAS.  However, with the Run 2 dataset a precise width measurement of the new resonance remained difficult (Figures~\ref{Resolution_ee_14TeV_300fb} and \ref{Resolution_mm_14TeV_300fb}).  There were also attempts to measure the forward-backward asymmetry ($A_{FB}$) of events in the Run 2 dataset at high-mass using the electron channel, to help differentiate between the various new physics models that predicted a $Z^{\prime}$ like resonance decaying to leptons.  The number of observed events in the current dataset meant that this was also difficult (Figure~\ref{Afb_ee_14TeV_300fb}).  Analysers attempted to interpret their $A_{FB}$ results using non-linear binning (Figure~\ref{Afb_ee_14TeV_300fb_Bin}) with the aim of achieving statistically significant model discrimination. Yet only the most widely varying models were able to be discriminated between when compared to the observed data. It was estimated that around an order of magnitude more data would be needed to start to make strong statements about the physics model that nature had presented the physicists with.

\subsection{Run 3 of the LHC}

\begin{figure}
\centering
\begin{minipage}{.5\textwidth}
  \centering
  \includegraphics[width=\linewidth]{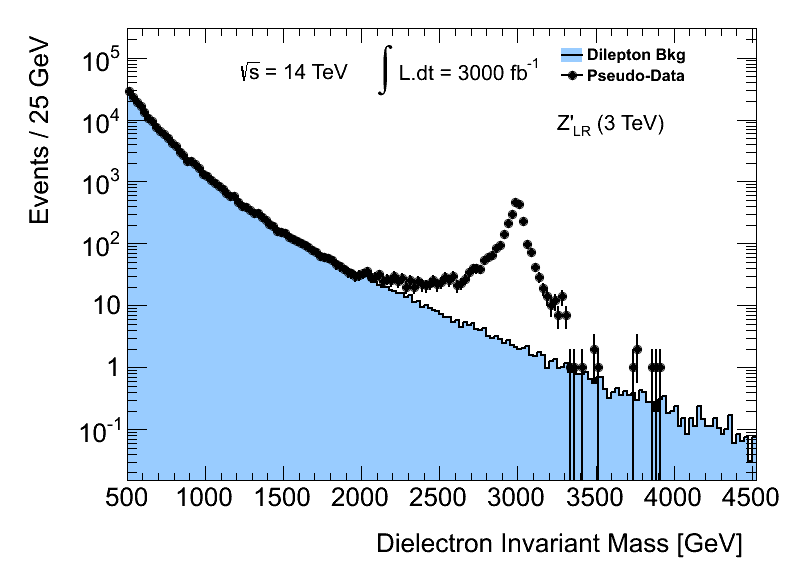}
  \caption{Dilepton backgrounds and the emerged signal for a LR $Z^{\prime}$ at 3~TeV for $e^+e^-$ pairs after 3000~fb$^{-1}$.}
  \label{SigPlusBkg_ee_14TeV_3000fb}
\end{minipage}%
\begin{minipage}{.5\textwidth}
  \centering
  \includegraphics[width=\linewidth]{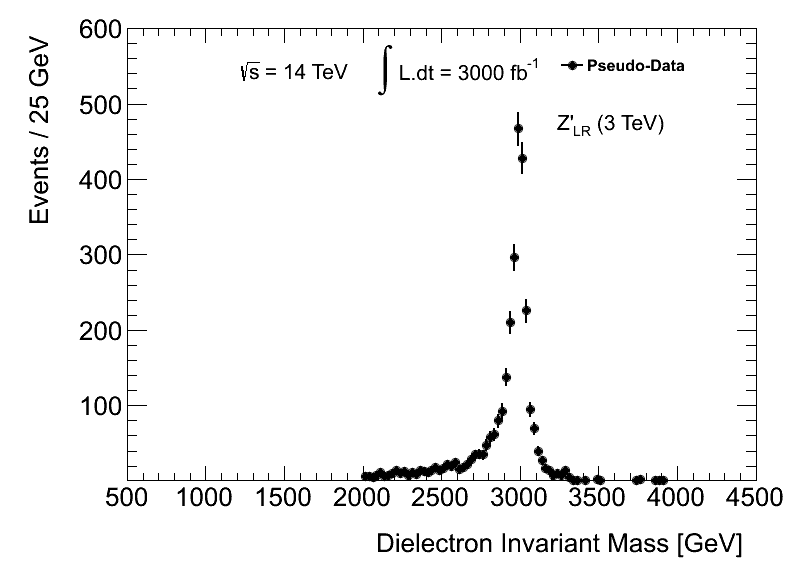}
  \caption{Emerged signal for a LR $Z^{\prime}$ at 3~TeV, background subtracted for $e^+e^-$ pairs after 3000~fb$^{-1}$.}
  \label{Sig_ee_14TeV_3000fb}
\end{minipage}
\end{figure}

\begin{figure}
\centering
\begin{minipage}{.5\textwidth}
  \centering
  \includegraphics[width=\linewidth]{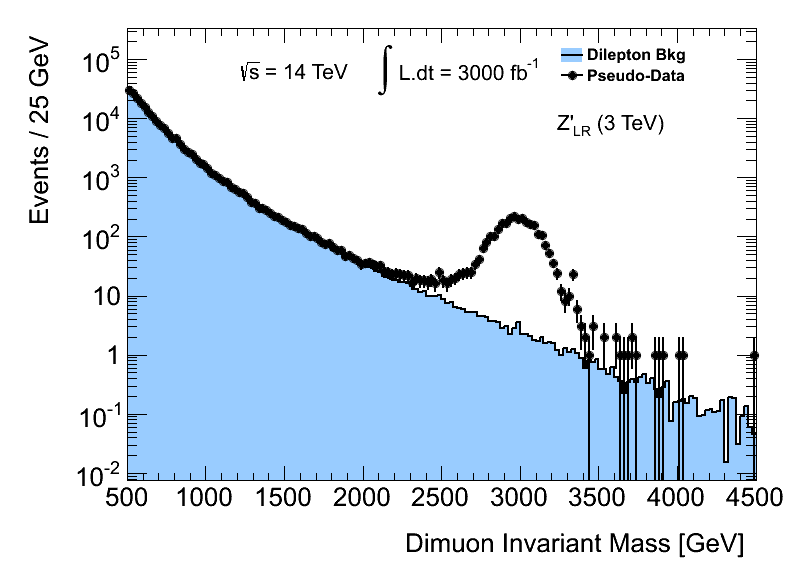}
  \caption{Dilepton backgrounds and the emerged signal for a LR $Z^{\prime}$ at 3~TeV for $\mu^+\mu^-$ pairs after 3000~fb$^{-1}$.}
  \label{SigPlusBkg_mm_14TeV_3000fb}
\end{minipage}%
\begin{minipage}{.5\textwidth}
  \centering
  \includegraphics[width=\linewidth]{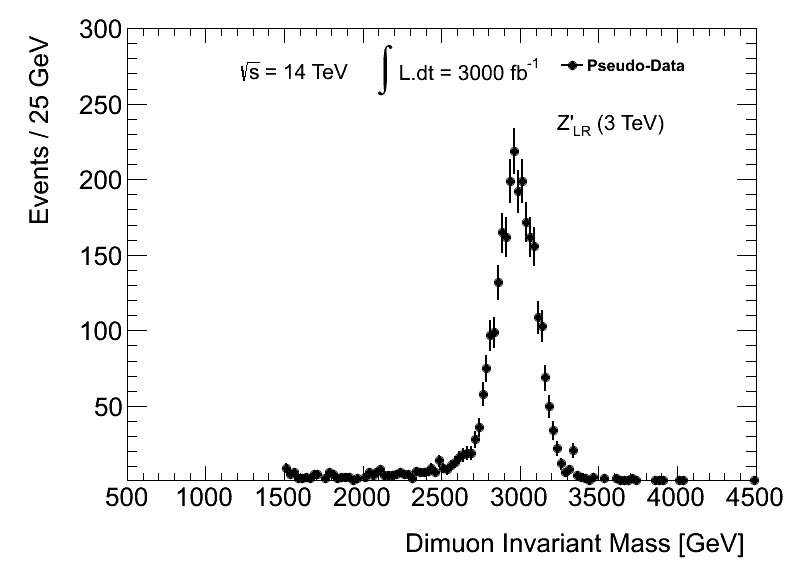}
  \caption{Emerged signal for a LR $Z^{\prime}$ at 3~TeV, background subtracted for $\mu^+\mu^-$ pairs after 3000~fb$^{-1}$.}
  \label{Sig_mm_14TeV_3000fb}
\end{minipage}
\end{figure}

\begin{figure}
\centering
\begin{minipage}{.5\textwidth}
  \centering
  \includegraphics[width=\linewidth]{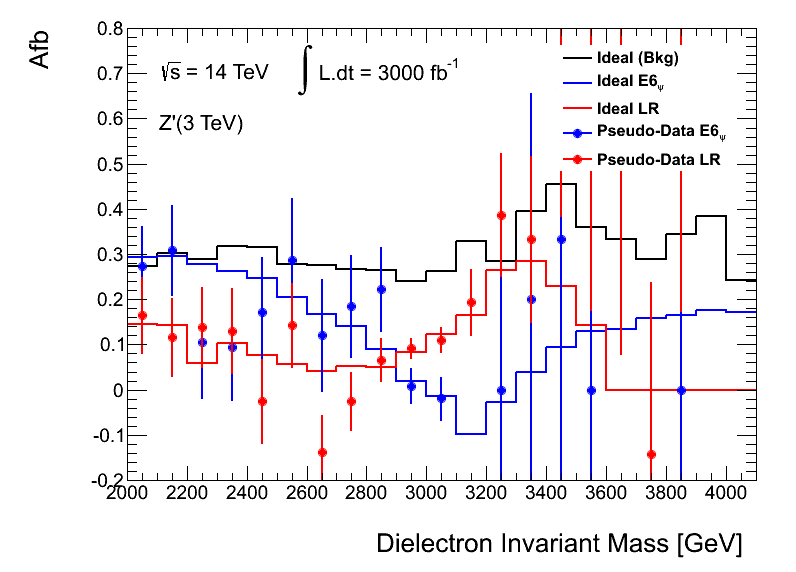}
  \caption{$A_{FB}$ of $e^+e^-$ pairs for the expected SM background (Black), as well as two signal scenarios for a 3~TeV resonance: $E_6$ model $Z^{\prime}_{\psi}$ (Blue), and LR model $Z^{\prime}_{LR}$ (Red). The solid lines show the ideal distributions, and colored data points show a single pseudo experiment after 3000~fb$^{-1}$.}
  \label{Afb_ee_14TeV_3000fb}
\end{minipage}%
\begin{minipage}{.5\textwidth}
  \centering
  \includegraphics[width=\linewidth]{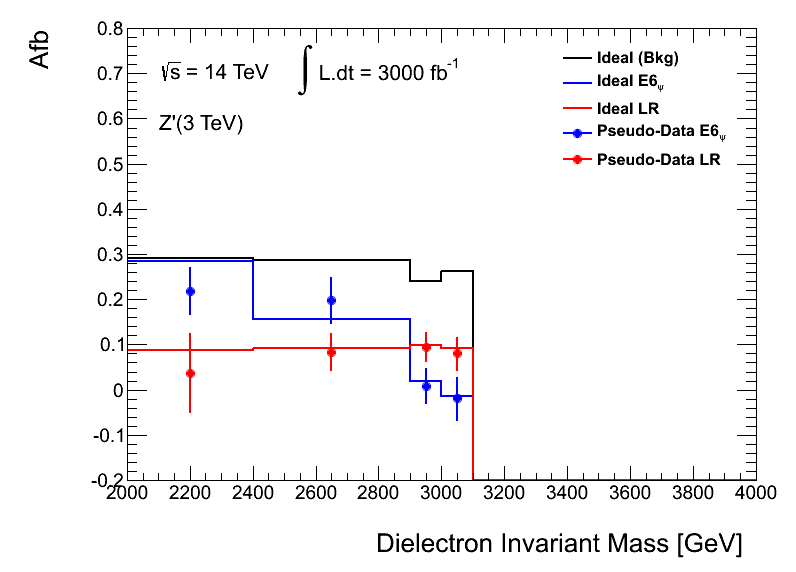}
  \caption{$A_{FB}$ of $e^+e^-$ pairs for the expected SM background (Black), as well as two signal scenarios for a 3~TeV resonance: $E_6$ model $Z^{\prime}_{\psi}$ (Blue), and LR model $Z^{\prime}_{LR}$ (Red). The solid lines show the ideal distributions, and colored data points show a single pseudo experiment after 3000~fb$^{-1}$. Here the tail statistics are combined into fewer bins.}
  \label{Afb_ee_14TeV_3000fb_Bin}
\end{minipage}
\end{figure}

Run 3 began on January 1, 2022 with the expectation of accumulating 3~ab$^{-1}$ over the next three years. The machine ran well and the Snowmass detector was able to contend with the pileup. After the run was ended, analysers now had the order of magnitude more data that they had estimated they needed to make statistically significant measurements of $A_{FB}$.  While the linear binning of this variable still proved difficult for model discrimination in all but the peak region, the non-linear binning of results confirmed over a good range, early hints of a LR model $Z^{\prime}$.  Invariant mass distributions are shown for both channels in Figures~\ref{SigPlusBkg_ee_14TeV_3000fb} to \ref{Sig_mm_14TeV_3000fb}, and $A_{FB}$ distribution shown for the electron channel in Figures~\ref{Afb_ee_14TeV_3000fb} and \ref{Afb_ee_14TeV_3000fb_Bin}.

\subsection{The $\sqrt{s}$ = 14~TeV Experiment Aftermath}

\begin{figure}
\centering
\begin{minipage}{.5\textwidth}
  \centering
  \includegraphics[width=\linewidth]{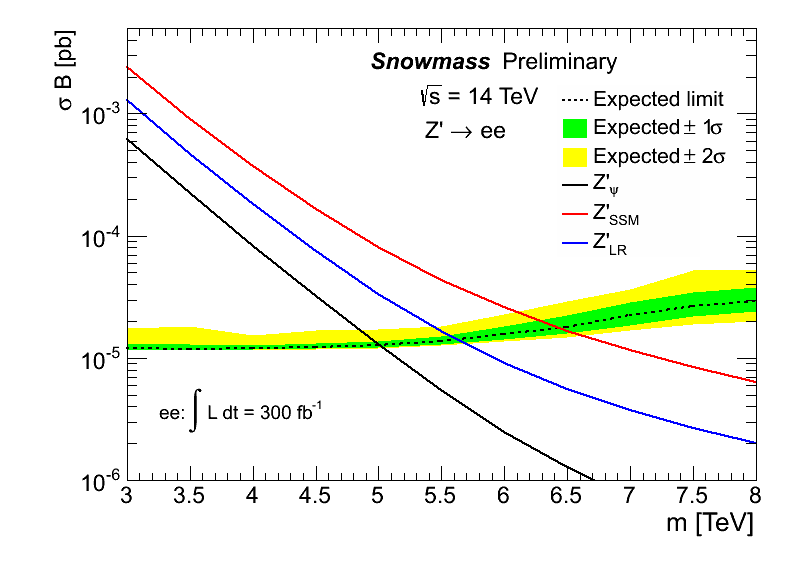}
  \caption{Upper cross-section limits for the process $q\bar{q} \rightarrow Z^{\prime} \rightarrow e^+e^-$, set at 95\% CL using a Bayesian statistical interpretation given 300~fb$^{-1}$ of data collected at $\sqrt{s}$ = 14~TeV.  Various signal scenarios are overlayed, with mass exclusion limits extracted at the intersection of the theory-expected lines.}
  \label{Limits_14TeV_300fb}
\end{minipage}%
\begin{minipage}{.5\textwidth}
  \centering
  \includegraphics[width=\linewidth]{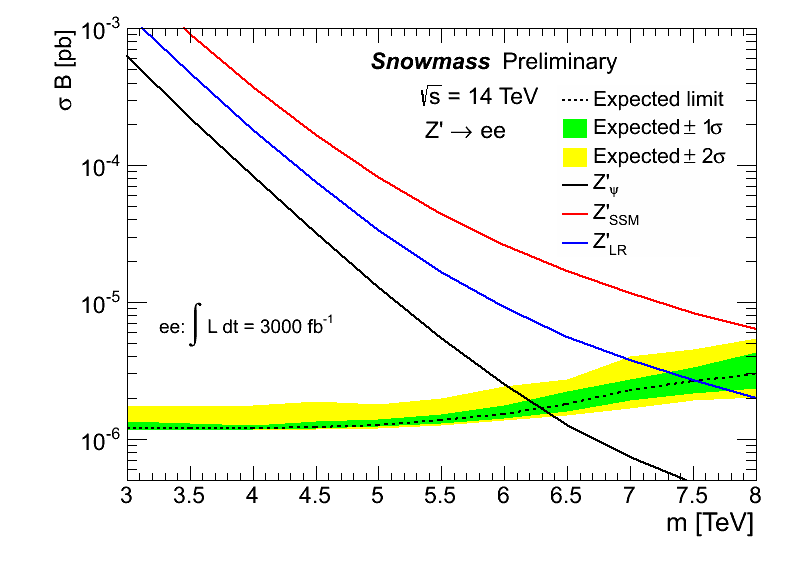}
  \caption{Upper cross-section limits for the process $q\bar{q} \rightarrow Z^{\prime} \rightarrow e^+e^-$, set at 95\% CL using a Bayesian statistical interpretation given 3000~fb$^{-1}$ of data collected at $\sqrt{s}$ = 14~TeV.  Various signal scenarios are overlayed, with mass exclusion limits extracted at the intersection of the theory-expected lines.}
  \label{Limits_14TeV_3000fb}
\end{minipage}
\end{figure}

With a total LHC dataset of 3~ab$^{-1}$ collected at $\sqrt{s}$ = 14~TeV, the discovery and initial measurements of a new $Z^{\prime}$ gauge boson at 3~TeV polemass had been made.  Should nature have presented a lighter $Z^{\prime}$, more events would have likely been produced and recorded, leading to better measurements.  Should nature have presented a heavier $Z^{\prime}$, most models would predict fewer events and thus model discrimination might not have been possible with this experiment, this story is just one scenario.  To imagine the range of possibilities, Figures~\ref{Limits_14TeV_300fb} and \ref{Limits_14TeV_3000fb} present the ability of the Snowmass experiment to exclude a $Z^{\prime}$ under different model assumptions, at $\sqrt{s}$ = 14~TeV and 300/3000~fb$^{-1}$ of collected data respectively.  The exclusion limits presented here are shown for the electron channel, but the muon channel would also give similar results, and the combination of channels giving marginally further reach still.  The resulting upper cross-section exclusion limits set at 95\% confidence level (CL) using a Bayesian statistical approach~\cite{BAT} with a flat positive prior for the signal cross-section time branching ratio ($\sigma$B) to leptons, is converted into a lower mass limit using the theoretical dependence of the signal cross-sections versus polemass, and presented in Table 1-1.

To reach further into the unknown, either increasing the production cross-section of any already newly discovered physics (such as a 3~TeV $Z^{\prime}$) to allow precision measurements, and/or push higher in our sensitivity to new physics at yet greater energetic regimes, the physicists appreciated a new (or greatly upgraded) accelerator would have to be built.

\begin{table}[htp]
\centering
\begin{tabular}{|c|c|c|c|}
\hline
$\int\cal{L}$.dt (at $\sqrt{s}$ = 14~TeV) & $Z^{\prime}_{\psi}$ [TeV] & $Z^{\prime}_{LR}$ [TeV] & $Z^{\prime}_{SSM}$ [TeV] \\
\hline
300~fb$^{-1}$ & 5.01 & 5.62 & 6.44 \\
3000~fb$^{-1}$ & 6.29 & 7.52 & $\sim$8.50 \\
\hline
\end{tabular}
\label{Numbers_Limits_14TeV}
\caption{Lower mass limits at 95\% CL for various $Z^{\prime}$ models given 300~fb$^{-1}$ and 3000~fb$^{-1}$ of collected data at $\sqrt{s}$ = 14~TeV, assuming no signal excess was observed.}
\end{table}

\section{The Discovery at a Future $\sqrt{s}$ = 33~TeV Experiment}

The end of the LHC Run 3 with its 14~TeV centre of mass energy and 3000~fb$^-1$ of collected data had come about in the year 2025.  Yet as far back at 2019, the Governments of the World had called upon the expertise of Engineers and Physicists to design and build a new machine that would probe further than the currently used accelerator.  One such possible proposal that could come to fruition was the High-Energy, High-Luminosity, Large Hadron Collider (HE-HL-LHC).  This machine would double the previous centre of mass energy to $\sqrt{s}$ = 33~TeV and aim to collect first 300~fb$^{-1}$ and then up to 3000~fb$^{-1}$ over two physics data taking Runs.  This machine would push the search reach for new physics to tens of TeV, truly surpassing anything before it and allowing any already newly discovered physics to bathe in increased production cross-sections and vast quantities of collected data that would allow some precision measurements, even though this was at a synchrotron machine.

\subsection{Run 1 of the Future Collider}

\begin{figure}
\centering
\begin{minipage}{.5\textwidth}
  \centering
  \includegraphics[width=\linewidth]{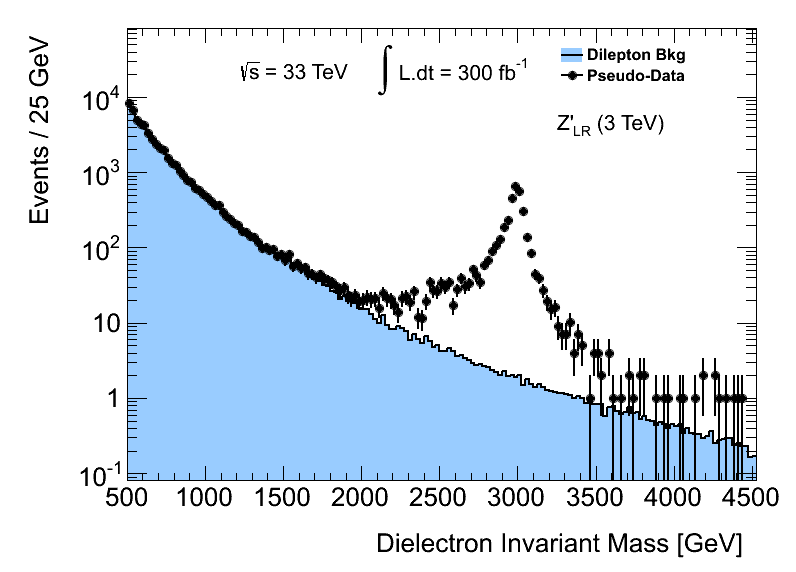}
  \caption{Dilepton backgrounds and the clear signal for a LR $Z^{prime}$ at 3~TeV for $e^+e^-$ pairs after 300~fb$^{-1}$.}
  \label{SigPlusBkg_ee_33TeV_300fb}
\end{minipage}%
\begin{minipage}{.5\textwidth}
  \centering
  \includegraphics[width=\linewidth]{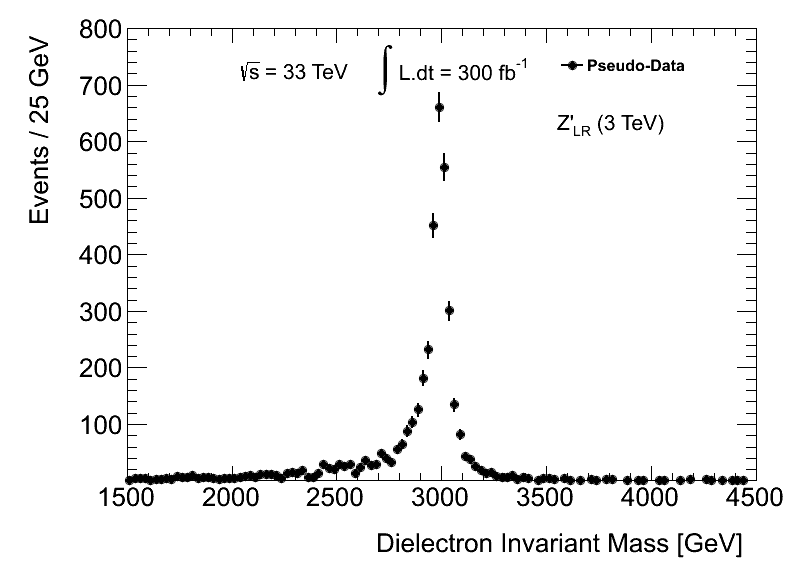}
  \caption{Fully emerged signal for a LR $Z^{\prime}$ at 3~TeV, background subtracted for $e^+e^-$ pairs after 300~fb$^{-1}$.}
  \label{Sig_ee_33TeV_300fb}
\end{minipage}
\end{figure}

\begin{figure}
\centering
\begin{minipage}{.5\textwidth}
  \centering
  \includegraphics[width=\linewidth]{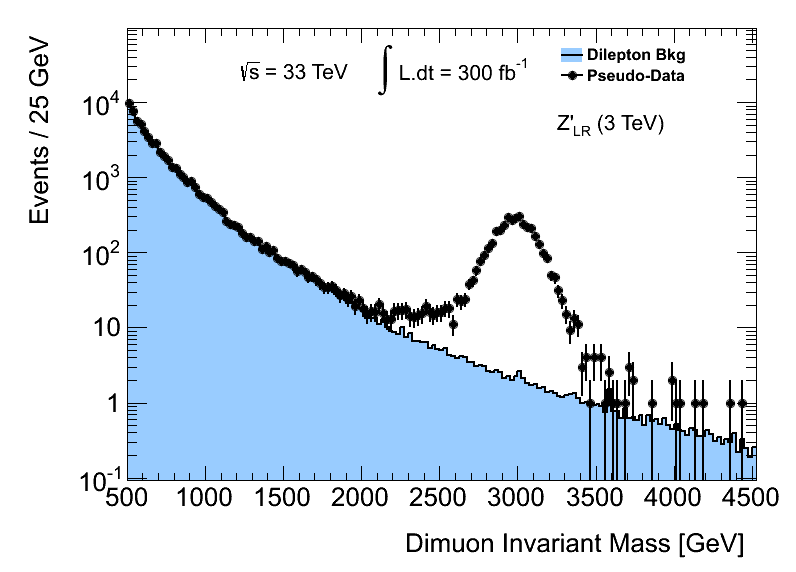}
  \caption{Dilepton backgrounds and the clear signal for a LR $Z^{\prime}$ at 3~TeV for $\mu^+\mu^-$ pairs after 300~fb$^{-1}$.}
  \label{SigPlusBkg_mm_33TeV_300fb}
\end{minipage}%
\begin{minipage}{.5\textwidth}
  \centering
  \includegraphics[width=\linewidth]{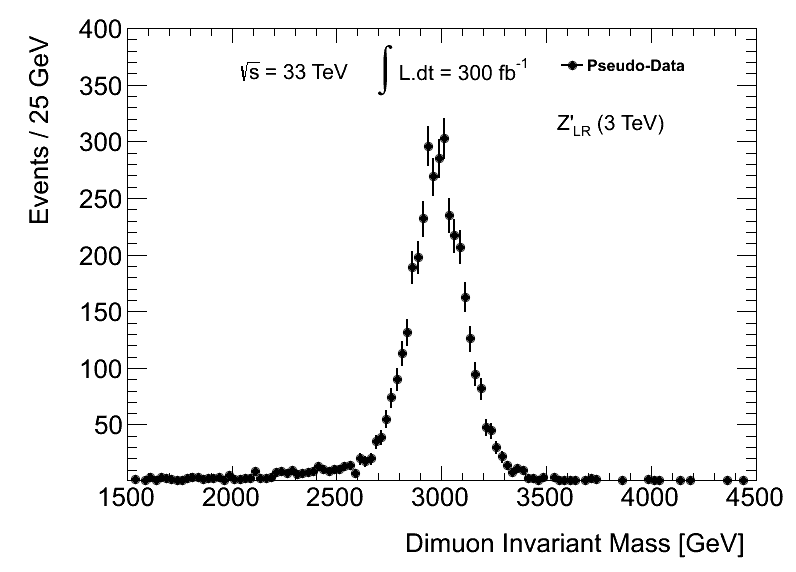}
  \caption{Fully emerged signal for a LR $Z^{\prime}$ at 3~TeV, background subtracted for $\mu^+\mu^-$ pairs after 300~fb$^{-1}$.}
  \label{Sig_mm_33TeV_300fb}
\end{minipage}
\end{figure}

\begin{figure}
\centering
\begin{minipage}{.5\textwidth}
  \centering
  \includegraphics[width=\linewidth]{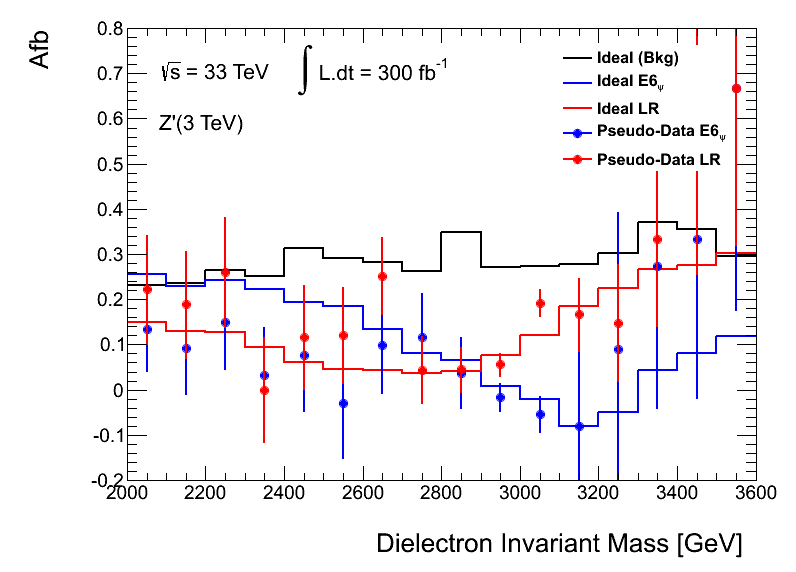}
  \caption{$A_{FB}$ of $e^+e^-$ pairs for the expected SM background (Black), as well as two signal scenarios for a 3~TeV resonance: $E_6$ model $Z^{\prime}_{\psi}$ (Blue), and LR model $Z^{\prime}_{LR}$ (Red). The solid lines show the ideal distributions, and colored data points show a single pseudo experiment after 300~fb$^{-1}$.}
  \label{Afb_ee_33TeV_300fb}
\end{minipage}%
\end{figure}

After the initial startup of the new accelerator in January 2026, at an astounding $\sqrt{s}$ = 33~TeV, the physicists and engineers quickly got used to their new experiment.  The Snowmass detector had been so expertly designed that a freshly built new version of the original was employed to collect data at this new experiment.  After three years of data taking with few problems or setbacks, the experiment had collected 300~fb$^{-1}$ of data, and many new stories in particle physics had unfolded.  Many physicists had continued to study one of the first newly discovered particles beyond the SM to be found at the previous experiment, namely the particle now relatively confidently called the Left-Right Symmetric Model (LRM) $Z^{\prime}$ gauge boson.  With 300~fb$^{-1}$ at $\sqrt{s}$ = 33~TeV, the experimenters were now already looking at a modestly increased number of confirmed $Z^{\prime}$ events, over the LHC Run 3 (Figures~\ref{SigPlusBkg_ee_33TeV_300fb} to \ref{Sig_mm_33TeV_300fb}), and reconfirming measurements they had made previously such as $A_{FB}$ with slightly increased precision (Figure~\ref{Afb_ee_33TeV_300fb}), some even combining the old and new datasets to get the most out of the data in hand.  They eagerly awaited the beginning of Run 2 in 2030 to push far beyond the experiences gained to date.

\subsection{Run 2 of the Future Collider}

\begin{figure}
\centering
\begin{minipage}{.5\textwidth}
  \centering
  \includegraphics[width=\linewidth]{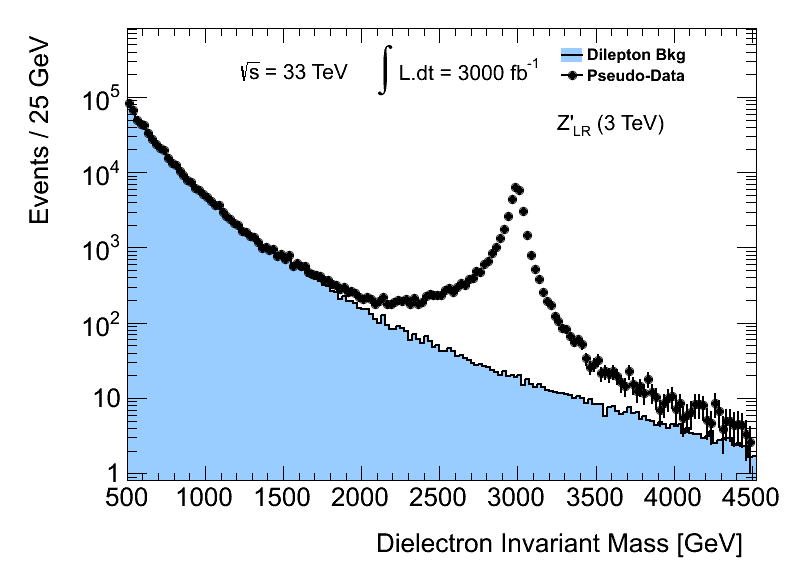}
  \caption{Dilepton backgrounds and the clear signal for a LR $Z^{\prime}$ at 3~TeV for $e^+e^-$ pairs after 3000~fb$^{-1}$.}
  \label{SigPlusBkg_ee_33TeV_3000fb}
\end{minipage}%
\begin{minipage}{.5\textwidth}
  \centering
  \includegraphics[width=\linewidth]{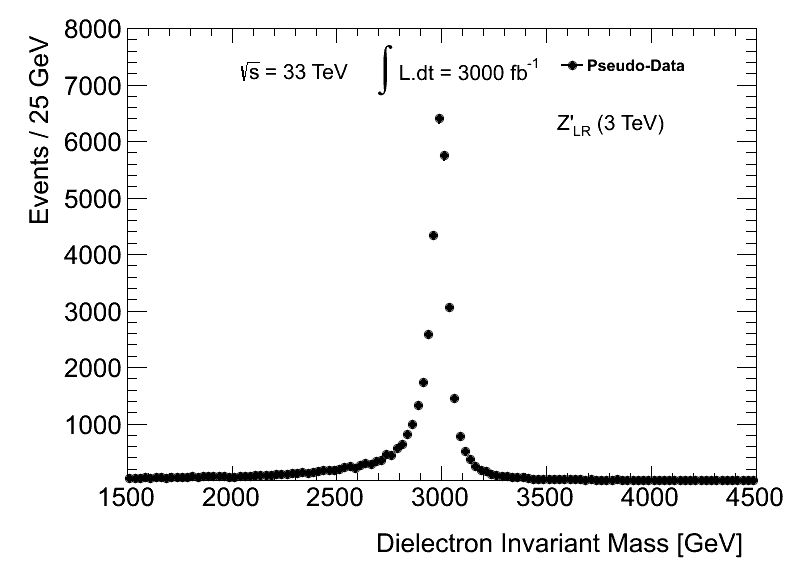}
  \caption{Fully emerged signal for a LR $Z^{\prime}$ at 3~TeV, background subtracted for $e^+e^-$ pairs after 3000~fb$^{-1}$.}
  \label{Sig_ee_33TeV_3000fb}
\end{minipage}
\end{figure}

\begin{figure}
\centering
\begin{minipage}{.5\textwidth}
  \centering
  \includegraphics[width=\linewidth]{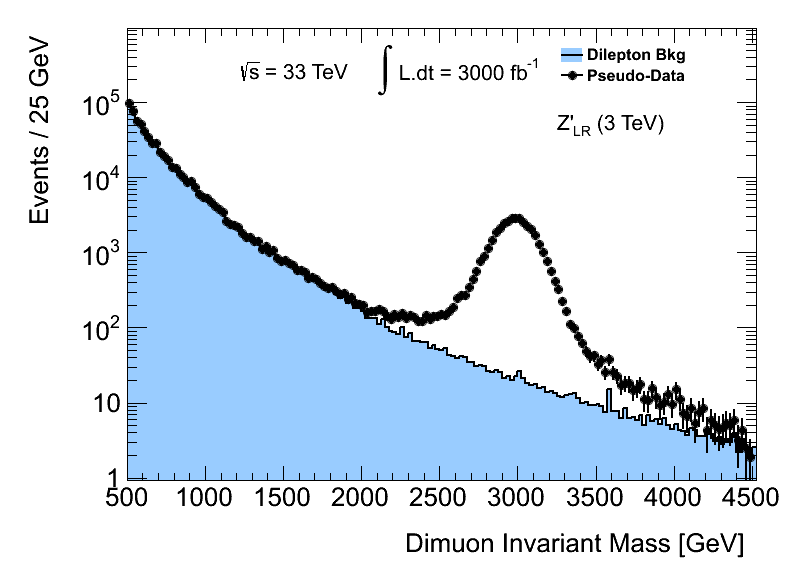}
  \caption{Dilepton backgrounds and the clear signal for a LR $Z^{\prime}$ at 3~TeV for $\mu^+\mu^-$ pairs after 3000~fb$^{-1}$.}
  \label{SigPlusBkg_mm_33TeV_3000fb}
\end{minipage}%
\begin{minipage}{.5\textwidth}
  \centering
  \includegraphics[width=\linewidth]{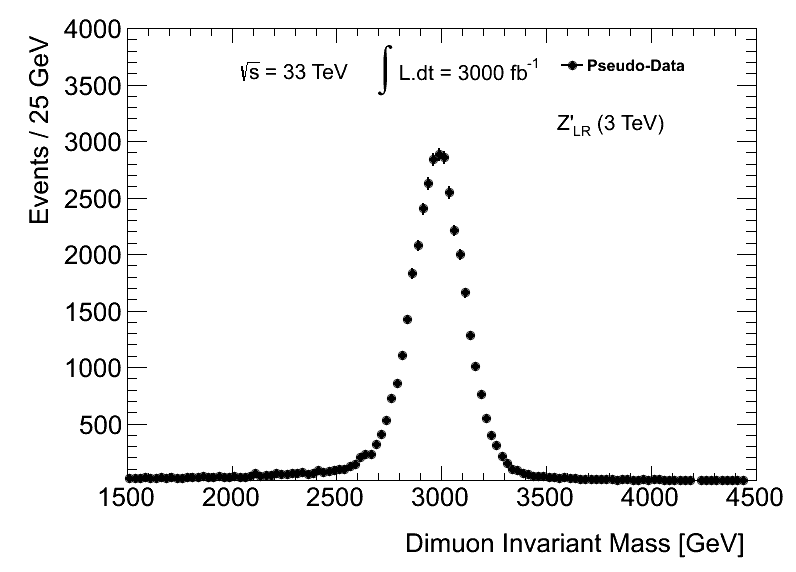}
  \caption{Fully emerged signal for a LR $Z^{\prime}$ at 3~TeV, background subtracted for $\mu^+\mu^-$ pairs after 3000~fb$^{-1}$.}
  \label{Sig_mm_33TeV_3000fb}
\end{minipage}
\end{figure}

\begin{figure}
\centering
\begin{minipage}{.5\textwidth}
  \centering
  \includegraphics[width=\linewidth]{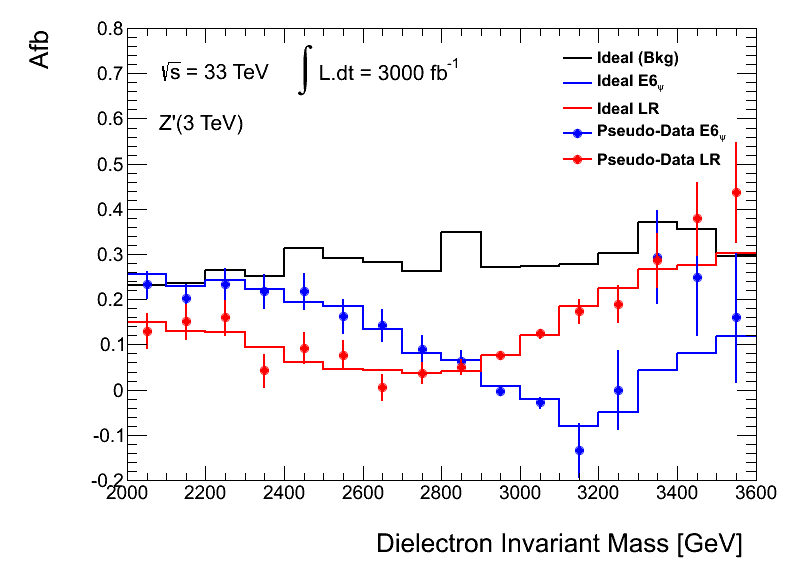}
  \caption{$A_{FB}$ of $e^+e^-$ pairs for the expected SM background (Black), as well as two signal scenarios for a 3~TeV resonance: $E_6$ model $Z^{\prime}_{\psi}$ (Blue), and LR model $Z^{\prime}_{LR}$ (Red). The solid lines show the ideal distributions, and colored data points show a single pseudo experiment after 3000~fb$^{-1}$.}
  \label{Afb_ee_33TeV_3000fb}
\end{minipage}%
\end{figure}

The beginning of Run 2 started in January of 2030 as expected without any delays.  Again, the data taking went smoothly, and other parallel stories of new physics continued to unfold as theorists struggled to simultaneously weave the numerous discoveries together into a new and over-arching tapestry explaining the fundamental laws of the Universe.  For the $Z^{\prime}$ story, tertiary measurements of SM couplings in specific decay channels and even the possible observation of exotic decays, were helping other stories understand their signal better as data was being recorded.  As run two ended in 2034, pile-up had continued to be a battle, but continually worked on and understood to bring an impressive dataset of 3000~fb$^{-1}$ at $\sqrt{s}$ = 33~TeV to the physics groups for analysis.  With this dataset the $Z^{\prime}$ analysis had been able to increase the number of recorded $Z^{\prime}$ events by an order of magnitude, bringing unprecendented levels of precision to measurements of width, mass, couplings, and even $A_{FB}$ (see complimentary white paper for in depth analysis~\cite{LianTao}).  The physicists remembered how far they had come from the first days of the LHC at $\sqrt{s}$ = 14~TeV, seeing a few events out at high-mass (Figure~\ref{SigPlusBkg_ee_14TeV_30fb}) and wondering if it would just turn out to be a fluctuation of the Standard Model.  Now the picture was very different, physicist's and indeed the World's understanding of the fundamental properties of the Universe had leaped almost unimaginably, and in the $Z^{\prime}$ analysis they were now presented with a magnificent and clear signal shape (Figures~\ref{SigPlusBkg_ee_33TeV_3000fb} to \ref{Sig_mm_33TeV_3000fb}), and $A_{FB}$ measurement that put the discovery of a LRM model $Z^{\prime}$ beyond all doubt (Figure~\ref{Afb_ee_33TeV_3000fb}). This new particle was one that they were almost getting used to, but which still excited even the newest Graduate students because of its implications and the theory paradigm shifts that had occurred over the last 15 years because of it.

\subsection{The $\sqrt{s}$ = 33~TeV Experiment Aftermath}

\begin{figure}
\centering
\begin{minipage}{.5\textwidth}
  \centering
  \includegraphics[width=\linewidth]{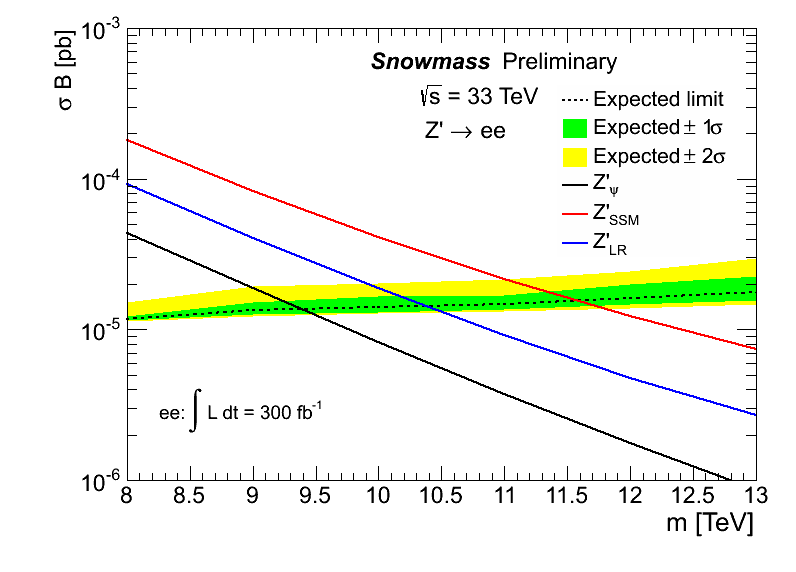}
  \caption{Upper cross-section limits for the process $q\bar{q} \rightarrow Z^{\prime} \rightarrow e^+e^-$, set at 95\% CL using a Bayesian statistical interpretation given 300~fb$^{-1}$ of data collected at $\sqrt{s}$ = 33~TeV.  Various signal scenarios are overlayed, with mass exclusion limits extracted at the intersection of the theory-expected lines.}
  \label{Limits_33TeV_300fb}
\end{minipage}%
\begin{minipage}{.5\textwidth}
  \centering
  \includegraphics[width=\linewidth]{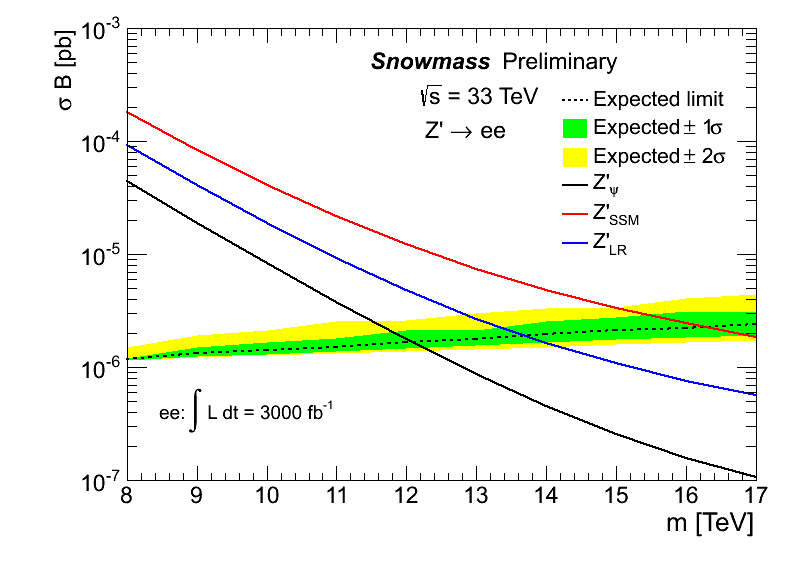}
  \caption{Upper cross-section limits for the process $q\bar{q} \rightarrow Z^{\prime} \rightarrow e^+e^-$, set at 95\% CL using a Bayesian statistical interpretation given 3000~fb$^{-1}$ of data collected at $\sqrt{s}$ = 33~TeV.  Various signal scenarios are overlayed, with mass exclusion limits extracted at the intersection of the theory-expected lines.}
  \label{Limits_33TeV_3000fb}
\end{minipage}
\end{figure}

The achievement of Engineers and Physicists alike was astounding, a new machine had been built to go up to energies of $\sqrt{s}$ = 33~TeV, and over 3000~fb$^-1$ of data had been collected from pp collisions over the years.  The journey was hard at times, and required continual maintenance and understanding of both the accelerator and the Snowmass detector, due to the incredibly harsh environment both were being subjected to, and the level of precision required for the physics analyses to thrive.
Again we break the fourth wall and note that the story played out here would be altered depending on what nature has in store, but that this story represents one credible path given the current exclusion limits set by LHC experiments at $\sqrt{s}$ = 8~TeV with 20~fb$^{-1}$ of data recorded in pp collisions.  To get an idea of the reach for new physics this imaginery machine would bring to the field, upper cross-section exclusion limits are set at 95\% CL for different $Z^{\prime}$ models, under the assumption of no observed excess.  These results are obtained using a Bayesian statistical interpretation with a flat positive prior for the signal $\sigma$B to leptons and presented in Figures~\ref{Limits_33TeV_300fb} and \ref{Limits_33TeV_3000fb} for 300/3000~fb$^{-1}$ respectively at $\sqrt{s}$ = 33~TeV.  The extracted lower mass limits for the various models are correspondingly shown in Table 1-2.
Whatever path of discovery for experimental particle physics lies ahead in the future, one thing is for certain: Every time a new frontier in the field is passed, new insights into nature are gained.  If and when a new particle is discovered, beyond our currently held description of nature, our understanding of the Universe will leap. This story has detailed from initial hints to final detailed analyses, how the particle physics community would set about investigating and understanding such a newly uncovered aspect of nature, which would likely be one of the first observable indications of new physics at a future machine.

\begin{table}[htp]
\centering
\begin{tabular}{|c|c|c|c|}
\hline
$\int\cal{L}$.dt (at $\sqrt{s}$ = 33~TeV) & $Z^{\prime}_{\psi}$ [TeV] & $Z^{\prime}_{LR}$ [TeV] & $Z^{\prime}_{SSM}$ [TeV] \\
\hline
300~fb$^{-1}$ & 9.39 & 10.37 & 11.57 \\
3000~fb$^{-1}$ & 12.01 & 13.70 & 16.26 \\
\hline
\end{tabular}
\label{Numbers_Limits_33TeV}
\caption{Lower mass limits at 95\% CL for various $Z^{\prime}$ models given 300~fb$^{-1}$ and 3000~fb$^{-1}$ of collected data at $\sqrt{s}$ = 33~TeV, assuming no signal excess was observed.}
\end{table}



\end{document}